\documentclass{article}
\usepackage{ijcai03}
\usepackage{times}
\usepackage{latexsym}
\usepackage{subfigure, epsfig}
\pdfoutput=1
\newif\ifpdf\ifx\pdfoutput\undefined\pdffalse\else\pdfoutput=1\pdftrue\fi

\title{Application of distributed constraint satisfaction problem \\
 to the agent-based planning in manufacturing systems}
\author{S. Kornienko, O. Kornienko, P. Levi \\
\small Institute of Parallel and Distributed High-Performance Systems, \\
\small University of Stuttgart, Universit{\"a}tsstr. 38, 70569
Stuttgart, Germany.
\thanks{\small Post-Proceeding Version 2004. Appeared in proceedings of the International Scientific Congress "Intelligent Systems (IEEE AIS'03)" and "Intelligent CAD's (CAD-2003)", p.124-140, Divnomorsk, Russia, 2003}}

\begin{document}
\maketitle

\begin{abstract}
Nowadays, a globalization of national markets requires developing flexible and demand-driven production systems. Agent-based technology, being distributed, flexible and autonomous is expected to provide a short-time reaction to disturbances and sudden changes of environment and allows satisfying the mentioned
requirements. The distributed constraint satisfaction approach underlying the suggested method is described by a modified Petri network providing both the conceptual notions and main details of implementation.
\end{abstract}

\section{Introduction}

An internationalization of market outlets, an individualization of consumer needs, a rapid implementation of technical innovation become a key factor of success in modern economic market. In order to survive in international competition, enterprises  are forced to react dynamically to new requirements, to make permanent modifications and adaptation of own structures. In particular this
concerns the planning processes.

A planning in manufacturing systems is traditionally organized
top-down. The strategical level of planning transmits the results
to the tactical level, which in turn triggers the operational
level of planning process. The final result of the planning
process is a detailed schedule of manufacturing to be implemented
on the shop floor. Every encountered disturbance, unforeseen in
the schedule, triggers a new planning cycle. Efforts, costs  and
time required for replanning can be essentially reduced by making
the planning process adaptive to disturbances. In this way a
modern flexible manufacturing requires a new approach that
introduces elements of self-organization into operational control.
Taking into account a spatial distribution of manufacturing
elements and requirement of flexibility to the whole system, the
concept of autonomous agents has found some applications in this
field \cite{Weiss}. This approach seems to be very promising to
guarantee the required robustness, fault tolerance, adaptability
and agility in the field of transformable manufacturing systems.

Application of agents to manufacturing requires also a development
of new and adaptation of known approaches towards typical problems
of multi-agents technology, like distributed problem solving,
planning or collective decision making \cite{Sandholm2}. This
paper deals with the distributed constraint-based short-term
planning (assignment) process supported by an multi-agent system (MAS).

\section{Assignment problem}
\label{s_assignment_problem}

The assignment problem is often encountered in manufacturing, it is a part of Operations Research / Management Science (OR/MS). It can be classified into scheduling, resources allocation and planning of operations order (e.g. \cite{Pinedo}). This is a classical $NP$-hard problem, there are known solutions by self-organization~\cite{Kornienko,Levi99}, combinatorial optimization \cite{Graves}, evolutionary approaches \cite{Blazewicz}, constraint satisfaction and optimization \cite{Alicke}, discrete dynamic programming \cite{Bellman}. However these methods are developed as central planning approaches, the distributed or multi-agents planning for the assignment problem is in fact not researched (overview e.g. in \cite{Durfee}).

Generally, this problem consists in assigning a lot of low-level jobs (like "to produce one piece with defined specification") to available machines so that all restrictions will be satisfied. The solution consists of four steps. The first step is to prove whether the machine is technological able to manufacture, whereas the second step is to
prove whether the machine is organizational available, e.g. it is without prearrangements. The first two steps formalize the constraint problem, where the following criteria should be taken into account \cite{KornienkoInd}:
\begin{itemize}
\item Technical criteria determine necessary features of a piece and on this basis it can be decided whether a machine is able to manufacture this kind of feature e.g. the drilling feature.
\item Technological criteria determine whether the machine can operate with a necessary quality or determine a technologically necessary order, e.g. the statutory tolerances.
\item Geometrical criteria result from a geometrical description of the workpiece, e.g. whether a necessary chucking is possible.
\item Organizational criteria are a set of specifications of production orders and machines. Firstly, it defines whether a machine has an available time slot and, secondly, whether this time slot is suitable to manufacture the given production order.
\item Optimization criteria like a cost or delivery time.
\end{itemize}

\begin{figure}[htp]
\centering 
\includegraphics[width=.25\textwidth]{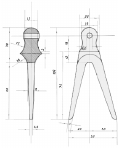}
\caption{\em \small Example of a workpiece to be manufactured.
\label{fig_b5_ws_of_ws}}
\end{figure}

All these criteria determine the agent-based process planning in the form of corresponding constraints, which reduce the decisions space in the assignment problem. The third and fourth steps of solution are the distributed constraint-satisfaction problem (CSP) and composition/optimization (COP) correspondingly. In the considered example it needs to manufacture
a multitude of parts and variants of production orders. Totalling
there are different types of workpieces (5-20 pieces of each type,
see Fig.~\ref{fig_b5_ws_of_ws})
that have to be manufactured on available machines.
Table~\ref{assignment2} shows an exemplary sequence of working steps,
where all mentioned technological
constraints are already considered.
\begin{table}[ht]
\centering
\begin{tabular}{cccccc}  \hline
WS   &  L/M 1 & L/M 2 & L/M 3 & Order   \\  \hline
  1         &     1      &     0     &     1     &    1    \\
  2         &     2      &     0     &     2     &    2    \\
  3         &     0      &     1     &     1     &    2    \\
  4         &     3      &     0     &     3     &    2    \\
  5         &     1      &     1     &     0     &    2    \\
  6         &     2      &     2     &     2     &    2    \\
  7         &     0      &     1     &     1     &    3    \\
  8         &     3      &     0     &     3     &    4    \\
  9         &     2      &     0     &     2     &    4    \\
  10        &     1      &     1     &     0     &    4    \\
  11        &     1      &     1     &     1     &    4    \\  \hline
\end{tabular}
\caption {\em \small Technological table for a workpiece, {\bf WS}
- working step, {\bf L/M} - length/machine. Zero in a length (of a
working step) at corresponding machine means that this machine
cannot perform the requested operation. Order of working steps
means, that e.g. the steps 2,3,4,5,6 should be produced after the
step 1 and before the step 7. It is natural to assume these steps
cannot be performed at the same time on different machines.}
\label{assignment2}
\end{table}

Processing of each workpiece consists of several working steps
(defined by a technological process), all these working steps
cannot be processed on one machine. Each from the working steps
has different length and also cost. Moreover each type of the
pieces has own technology, i.e. a processing consists of different
working steps. For simplification it is assumed that available
machines are of the same type, therefore the cost and length of
the same working step do not differ on these machines (in general
case they are different). The aim is to generate a plan of how to
manufacture these workpieces with minimal cost, minimal time (or
other optimization criteria), taking into account restrictions
summarized in Table~\ref{assignment2}.

\begin{figure}[htp]
\centering
\includegraphics[width=.45\textwidth]{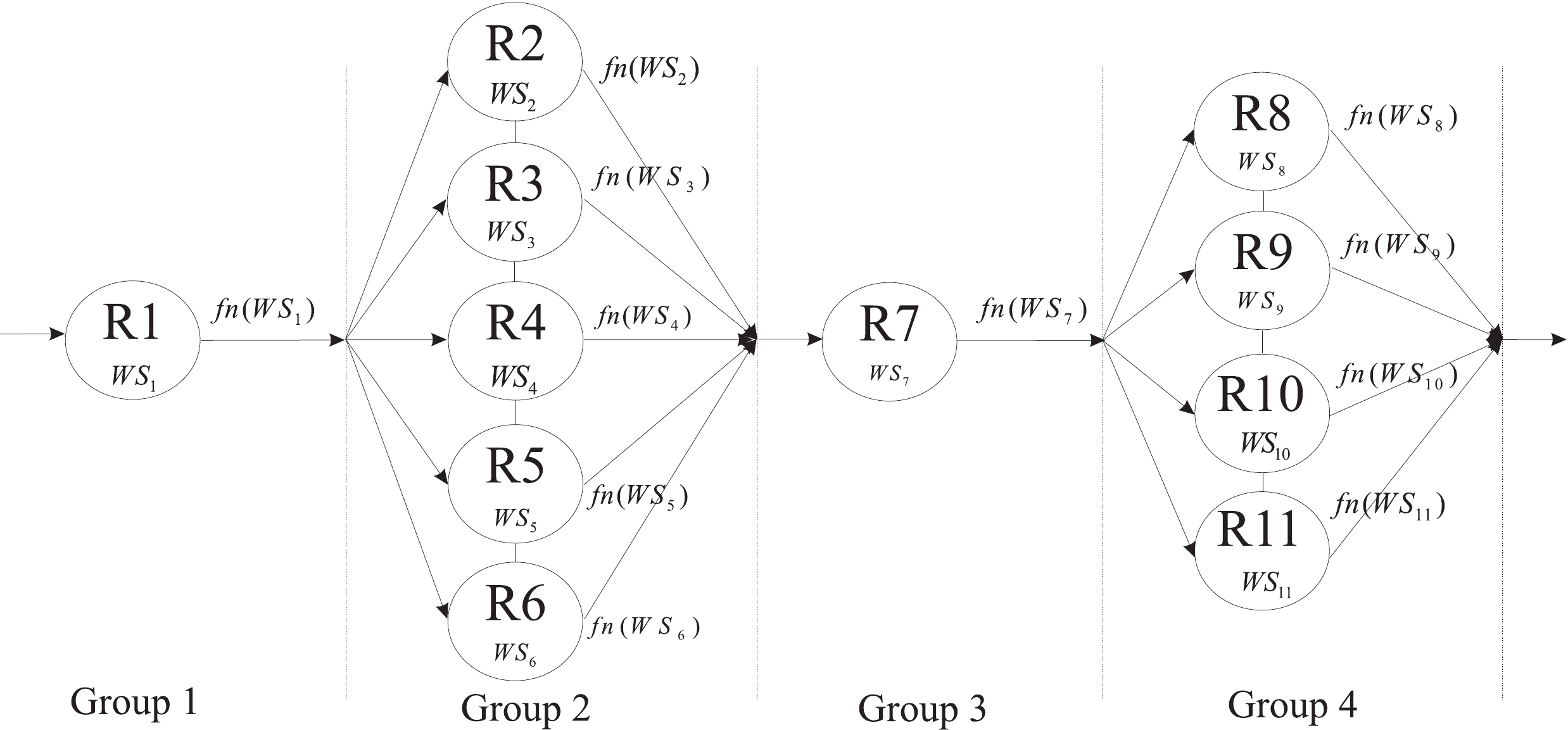}
\caption{\em \small Constraint network for the assignment problem.
\label{constraint}}
\end{figure}

Let us denote a working step as $WS^i_j$, where $i$ is a type of workpieces and $j$ a number of the working step, an available machine is denoted as $M_k$, where $k$ is a number of machine. We need also to introduce a piece $P^m_n$, where $m$ is a priority of production and $n$ is a number of this piece. In this way $st(P^m_n(WS^i_j))$, $fn(P^m_n(WS^i_j))$ denote a start and a final positions of the corresponding working step that belongs to the corresponding
piece ($st(WS^i_j)$, $fn(WS^i_j)$ for all pieces). We start with the definition of these values
\begin{eqnarray*}
P^{m \in [1-20] }_{n \in [1-20]} (WS_{j\in  [1,...,11 ] }^{i\in \{ A,B,C,D,E \} })&=&o \in operation, \\
M_{k \in \{ 1,2,3,4 \} }&=& \{o \in  operation \}, \\
st(P^m_n(WS^i_j))&=&\{ t\ge 0, t \in R \}, \\
fn(P^m_n(WS^i_j))&=&\{st(P^m_n(WS^i_j))+\\&+&length(P^m_n(WS^i_j)) \}.
\end{eqnarray*}
The first constraint determines a correspondence between operations of working step and of the $k$-machine
\begin{eqnarray*}
C_1 = \{(o_1,o_2)|o_1 \in P^m_n(WS^i_j), o_2 \in M_k, o_1=o_2 \}.
\end{eqnarray*}
The technological restrictions given by the Table \ref{assignment2} can be
rewritten in the following form
\begin{eqnarray*}
C_{2} &=& \{(fn(WS^A_{[1]}) < st(WS^A_{[2-6]})) \subset WS^A_j \times WS^A_j\}, \\
C_{3} &=& \{(fn(WS^A_{[2-6]}) < st(WS^A_{[7]})) \subset WS^A_j \times WS^A_j\},  \\
C_{4} &=& \{(fn(WS^A_{[7]}) < st(WS^A_{[8-11]})) \subset WS^A_j \times WS^A_j\},
\end{eqnarray*}
where $WS^A_{[2-6]}$ (for all pieces) cannot be performed at the same time
\begin{eqnarray*}
C_5 &= &\{(j \in  [st(WS^A_{[w]}),...,fn(WS^A_{[w]})] \neq  \\
&\neq j& \in [st(WS^A_{[w']}),...,fn(WS^A_{[w']})]) \subset
\nonumber \\
& \subset& WS^A_j \times WS^A_j; w,w'=2,3,4,5,6; w \neq w' \}
\end{eqnarray*}
and also $WS^A_{[8-11]}$
\begin{eqnarray*}
C_6&=& \{(j \in  [st(WS^A_{[w]}),...,fn(WS^A_{[w]})] \neq   \\
&\neq& j \in [st(WS^A_{[w']}),...,fn(WS^A_{[w']})])
\nonumber \\
& \subset& WS^A_j \times WS^A_j; w,w'=8,9,10,11; w \neq w'\}.
\end{eqnarray*}
Priority of production can be expressed by
\begin{eqnarray*}
C_7 &=&  \{(m \in P^{m}_n > m \in P^{m}_{n'} | st(P^m_n(WS^A_{j}))
> \\
&>&  st(P^m_{n'}(WS^A_{j}))) \subset P^{m}_n \times P^{m}_n, n
\neq n' \}.
\end{eqnarray*}
As soon as the variable, the domains of values and constraints are defined, a propagation approach can be started. The aim is to restrict the values of variables (or to find such values of variables) that will satisfy all constraints. This propagation can be represented in the way shown in Fig.~\ref{constraint}.

All working steps that belong to the same workpiece build a sequence. Every node in this sequence gets a "finish"-position of a working step from the previous node. Using this value, a current node looks for "start"-positions of the next working step that satisfy local constraints, calculates "finish"-positions and propagates them to the next node. If no position satisfying local constraint can be found, the node
requests another "finish"-position from the previous node. In this way the network can determine locally consistent positions of all working steps. After that the obtained values should be tested for a global consistence.

\section{Application of the autonomous agents}
\label{s_application}

The CSP approach described in the previous section is necessary and sufficient in solving the discussed kind of assignment problem. Being implemented by one of programming techniques, it will generate the required plan. However working in a presence of disturbances (like machine failure, technological change and so on) requires additional efforts to adapt the planning approach to these changes. The principles of such an adaptation are not contained in the plan itself, an additional mechanism is needed. As mentioned in the introduction, the multi-agent concept can be used as such a mechanism that lends to manufacturing system more flexibility to
adapt to disturbances. But what is the cause and cost of this additional feature~? There is a long discussion of this point based e.g. on the decentralization (e.g. \cite{Decentralization}) or several dynamics properties of MAS (e.g in \cite{Weiss}). We would like to add the following argument into this discussion.

The multi-agent system can be considered from a viewpoint of theory of finite-state automata. Transition of $m$-states automaton (with or without memory, it does not change the matter) from one state to another is determined by some rules (by a program), therefore the automaton behaves in completely deterministic way. If a
control cycle is closed (see e.g. \cite{Weiss}) the automaton is autonomous, i.e. behaves independently of environment (other automata). Now consider a few such automata coupled into a system {\it in the way that keeps their autonomy}.
Forasmuch as each automaton behaves according to own rules, there is no a central program that determines a states transition of the whole system. In the "worst case" coupling $n$ automatons with $m$ states, the coupled system can demonstrate $n^m$ states.

Evidently this "worst case" has never to arise in the system, but how to control a behaviour of the distributed system without a central program (without a centralized mediator)~? The point is that all automata are continuously communicating in order to synchronize own states in regard to environment, to the solving task, etc. (in this case the notion of an automaton is replaced by the notion of an agent). The agents during communication "consider" all possible states and then "choose" such a state that is the most suitable to a currently solving problem. This is a main difference to the "centralized programming" approach. The central program can react only in such a way that was preprogrammed. For example 10 agents with 10 states can demonstrate $10^{10}$ different combinations. However no programmer is able to predict all situations to use all these states. Thus, the "centralized programming"
approach restricts the multi-agent system although there are essentially more abilities to react.

The sufficient number of degrees of freedom represents a key problem of multi-agent technology. On the one hand if the system is hard restricted in the behaviour, the advantage of MAS is lost. On the other hand, if the system has too many degrees of freedom it can communicate an infinitely long time. In other words
only several combinations of agents states have a sense and the point is how to achieve and to manage these states. This is a hard problem arising in many branches of science and engineering and correspondingly there are several
ways to solve it. The suggested here solution is based on a hierarchic software architecture that supports agent's autonomy.

\begin{figure}[htp]
\centering 
\includegraphics[width=.3\textwidth]{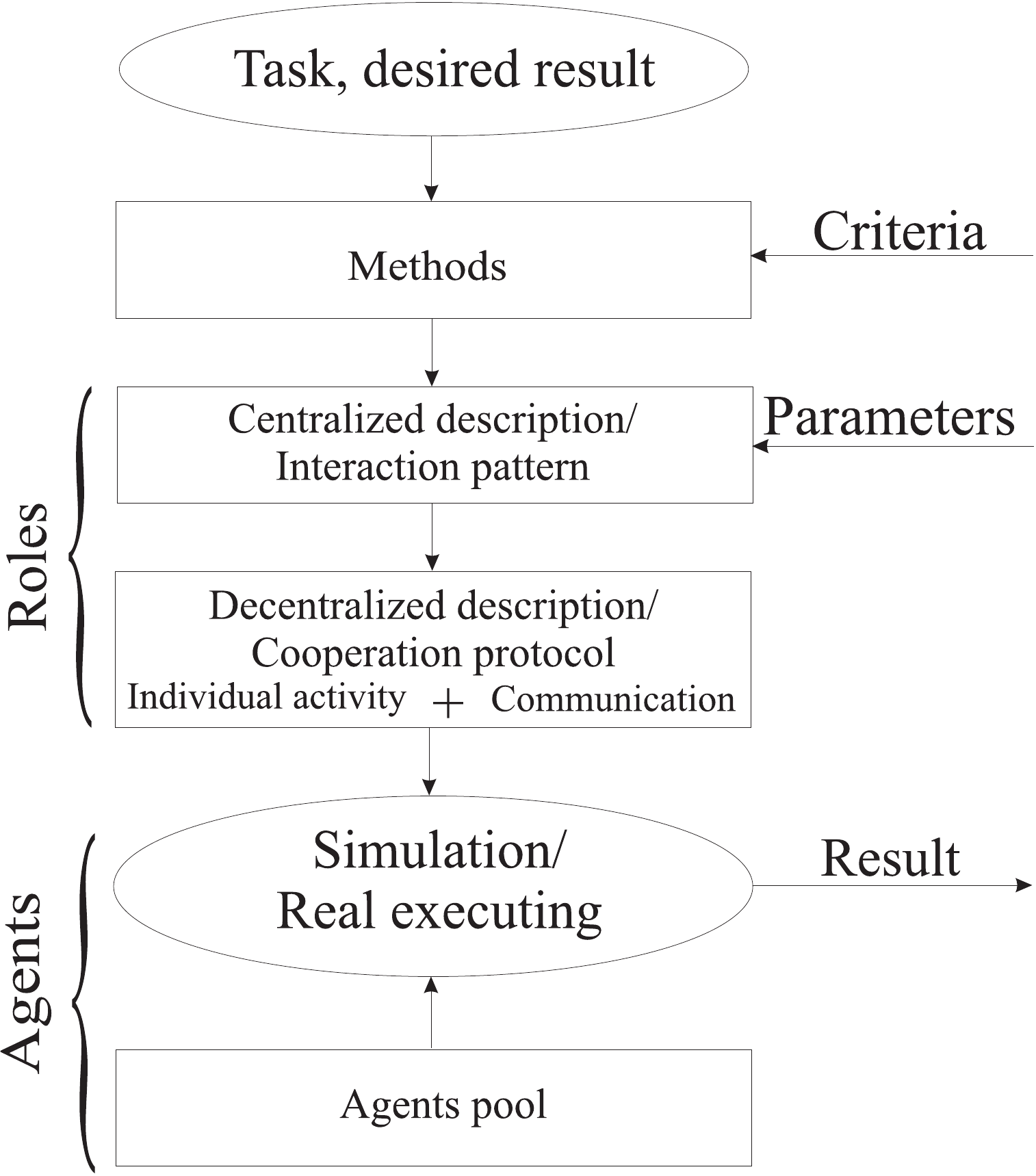}
\caption{\em \small Methodological approach towards agent-based applications.
\label{general_ideas}}
\end{figure}

Before starting to describe an approach, it needs to mention one methodological point concerning decentralization of multi-agent system, shown in Fig.~\ref{general_ideas}.
The MAS solves a problem by using  some methodological basis. For example the CSP and COP approaches basically underlie the solution of constraint problems. The point is that a methodological basis, in almost all cases, is formulated in a centralized way. It looks like a "battle plan", where all agents and their interactions are shown. Therefore this global description is often denoted an {\it interaction pattern}.

However the agents do not possess such a global point of view and the interaction pattern has to be distributed among agents. This decentralization concerns global information, messages transfer, synchronization, decision making and so on. The decentralized description of the chosen method should determine an individual activity of an agent as well as its interaction with other agents. It is also important that all agents behave in ordered way, i.e. to include cooperation mechanisms (protocols) into this distributed description~\cite{Constantinescu04}. In order to enable a transition from the interaction pattern to the {\it cooperation protocol} (see Fig.~\ref{general_ideas}) a notion of a role is introduced~\cite{ROPE}. A role is associated with a specific activity needed to be performed (according to a methodological basis). Agent can "play" one role or a sequence of roles. In this way interactions are primarily determined between roles, an agent (with corresponding abilities) handles according to the role playing at the moment. An advantage of this approach is that the centralized description (familiar for human thinking) is preserved whereas the roles in the interaction pattern are "in fact" already distributed, i.e. a mapping "agent-on-role" may be performed in a formalized way by a program. Thus, an interaction pattern is a "mosaic image" that from afar looks like a common picture (method), but at a short distance is a set of separate fragments (roles). Moreover a concept of roles allows decoupling the structure of cooperation processes from agents organization, so that any modification of agents does not affect the cooperation process and vice versa \cite{ROPE}.

The interaction pattern determines a {\it primary activity} (primary algorithm) of multi-agent system~\cite{Kornienko_S04}. The primary algorithm includes also some parameters whose modifications can be commonly associated with disturbances. Variation of these parameters does not disturb an activity of agents. In this case these are the expected disturbances, a reaction of the system on them is
incorporated into the primary algorithm. However due to specific disturbances every agent can reach such a state that is not described by a primary algorithm and where a performing of the next step is not possible. In this case the agent is in emergency
state and tries to resolve the arisen situation all alone  or with an assistance of neighbour agents ({\it secondary activity}). If the abilities of an agent are not sufficient or it requires additional resources it calls a rescue agent. The rescue agent is an agent that possesses specific (usually hardware) abilities. Anyway, the aim of agents in emergency state is to change a part of the primary algorithm so that to adapt it to  disturbances. The disturbances causing local emergency are expected (predicted) but not introduced into the primary algorithm.

The primary algorithm as well as its parameterization is optimal only for specific conditions (e.g. combinatorial/heuristic methods for solutions of combinatorial problems, CSP/COP for constraints, etc.). If disturbances change these conditions the primary algorithm became non optimal and it has no sense to repair it. All
agents have collectively to recognize such a global change and to make a collective decision towards replacement of the primary algorithm. This change corresponds to a global emergency. The disturbances causing the global emergency are not expected
(predicted), however they influence the conditions of primary algorithms and in this way can be recognized. Finally, there are such disturbances that cannot be absorbed by any changing of an algorithm, they remain irresolvable. Forasmuch as all these points cannot be considered in details in the framework of the paper, we restrict ourselves only by treating the mentioned primary algorithm in the language of cooperative processes by the modified Petri nets. The secondary activity as well as emergency states and rescue agents are not considered here.

\subsection{The primary algorithms}

In the case of an assignment planning, the primary algorithm is determined by the CSP approach described in Sec.~\ref{s_assignment_problem} (generally usage of
constraint-based approaches for MAS is not new, see e.g.~\cite{Nareyek}). Each working step in the approach is represented by a node in the constraint network shown in Fig.~\ref{constraint}. These nodes are separated from one another,
moreover their behaviour is determined by propagations. Therefore it is natural to give a separate role (R) to each node. However, before starting a propagation, this network has to be created and parameterized by technology, machines, number of workpieces and so on. These two steps (parameterization and propagation) will be
described by interaction patterns using corresponding roles.

\begin{figure}[htp]
\centering
\includegraphics[width=.4\textwidth]{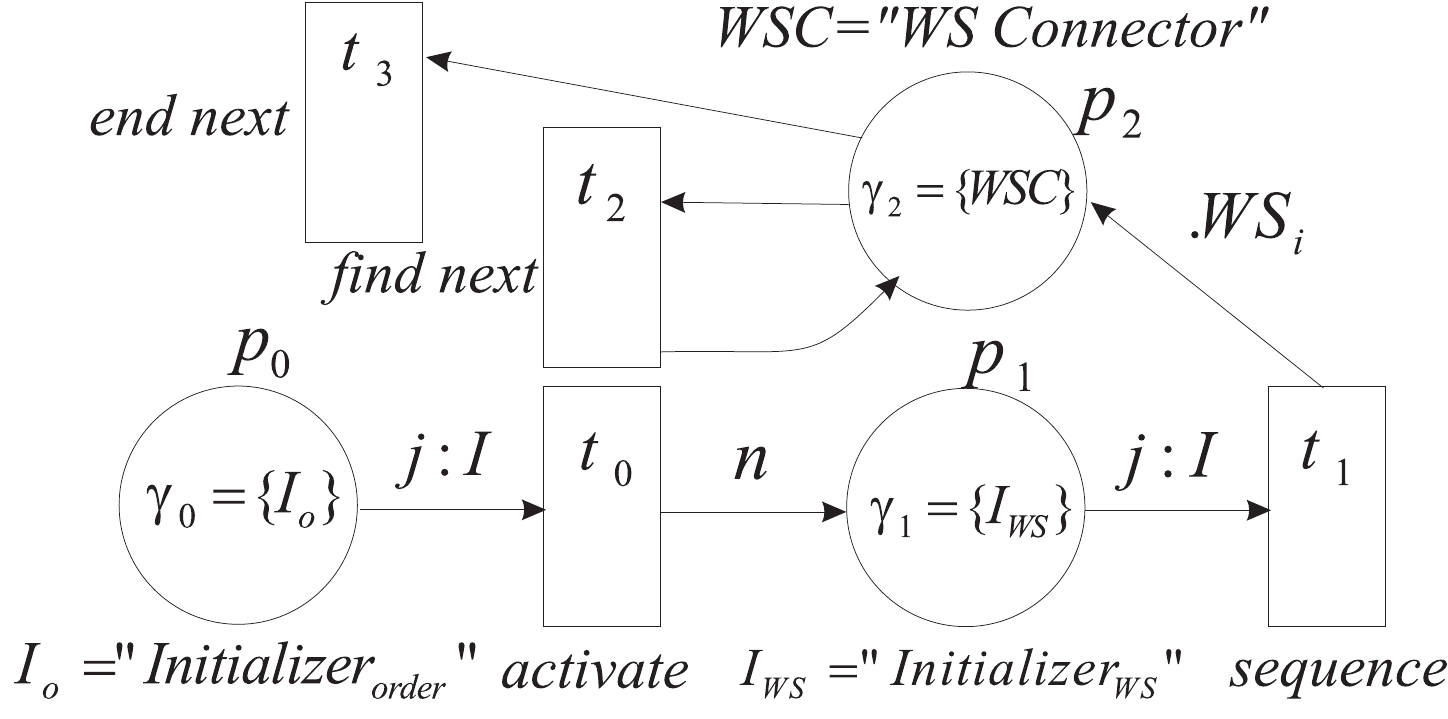}
\caption{\em \small Primary algorithm: Parameterization part.
\label{cooparation_start}}
\end{figure}

For description of agent's activities the RoPE (Role oriented Programming Environment) methodology is developed. Using this approach, the roles executed by agents are described in a formal way by the Perti network. Details of the RoPE system as well as description of cooperations protocols via Perti networks can be
found e.g. in~\cite{KornienkoInd},~\cite{Kornienko_S03A}. As already mentioned, the primary algorithm consists of two parts, parameterization and propagation, that represent a linear sequence of activities. The parameterization part, shown in
Fig.~\ref{cooparation_start}
has three phases $p_0$,  $p_1$, $p_2$ whose main result consists in determining a structure, neighbourhood relations and parameterization of nodes of the constraint network. The roles $\gamma_0, \gamma_1$ are "Initializers"
of WS-order and WS-nodes correspondingly. The role $\gamma_0$ is activated by the first production order, this role reads resource-objects and determines how much nodes (WS-roles) are required. The transition $t_0$ proves whether the result of $j.returnEND()$ is true (action is successful) and activates $\gamma_1$ with parameter $n$ as a number of required nodes. The $\gamma_1$ initializes each node according to all restrictions (technology, propagation rules, number of machines and so on). If this activity is also successful (transition $t_1$), the third role $\gamma_2$ is activated. It connects the created nodes (return a pointer to previous node), composing in this way a network. This interaction plan is finished (transition $t_3$) if there exists no next node needed to be connected.

\begin{figure}[htp]
\centering
\includegraphics[width=.4\textwidth]{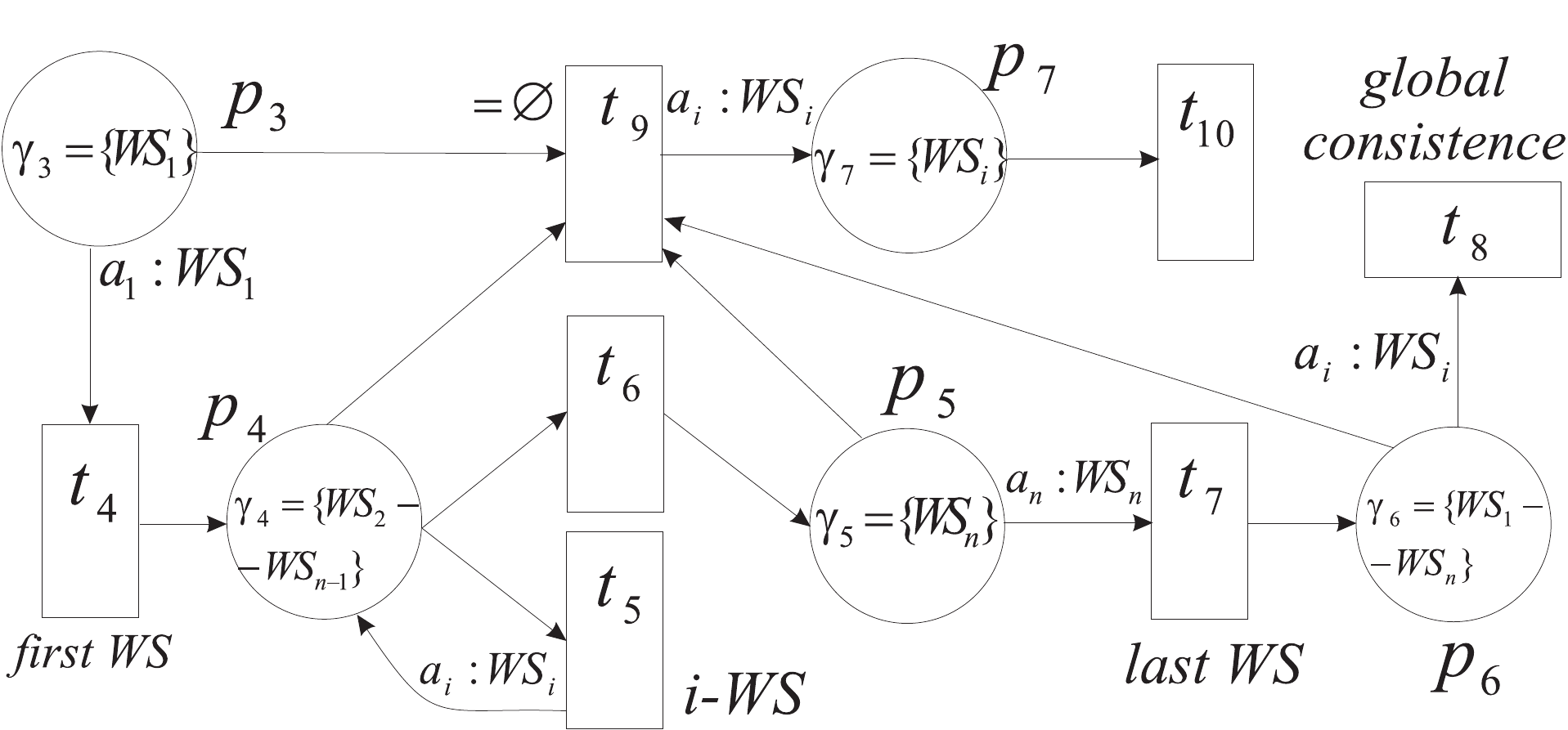}
\caption{\em \small Primary algorithm: Propagation part.
\label{cooperatipo_propagation}}
\end{figure}

The propagation part, shown in Fig.~\ref{cooperatipo_propagation},
consists of three blocks: local (the phases $p_3,p_4,p_5$), global (the phase $p_6$) propagations and an activity (the phase $p_7$) in the case of empty sets. The roles $\gamma_3, \gamma_5$ determine the propagation in the first
and last nodes whereas $\gamma_4$ does the same for all other nodes. The transition $t_7$ proves whether the local propagation was successful for all nodes and activates then the global propagation in $\gamma_6$. We emphasize the
local propagation requires a sequential executing of roles whereas in global propagation all roles can be in parallel executed. Finally, the transition $t_9$ proves whether the values set (WS-positions) of each node is empty. In the case of empty sets the role $\gamma_7$ tries to increase initial areas of values, first locally in neighbour nodes, then globally by restart of the local propagation.

Thus, agents, executing the roles described in the parameterization and propagation parts, "know how to solve" the CSP problem. Therefore all disturbances associated with a change of resources and constraints can be absorbed by a change of
parameters in the interaction patterns. In this way the MAS has enough degrees of freedom to be adaptive to disturbances in a framework of the primary algorithm.

\section{Agent-based optimization}
\label{s_optimization}

The steps described in the previous sections allow generating the sequence of working steps that satisfies all local constraints. However such global characteristics of a plan as cost, manufacturing time and so on are not considered. Therefore, as pointed out by some authors, the next step consists in optimizing the obtained
sequences. Generally speaking, constraint satisfaction and optimization cannot be separated into two different steps, rather, it represents a sui generis combination. Before to start a discussion of agent-based optimization it needed to mention two features of such an approach.

\begin{figure}[htp]
\centering
\subfigure[]{\includegraphics[width=.23\textwidth]{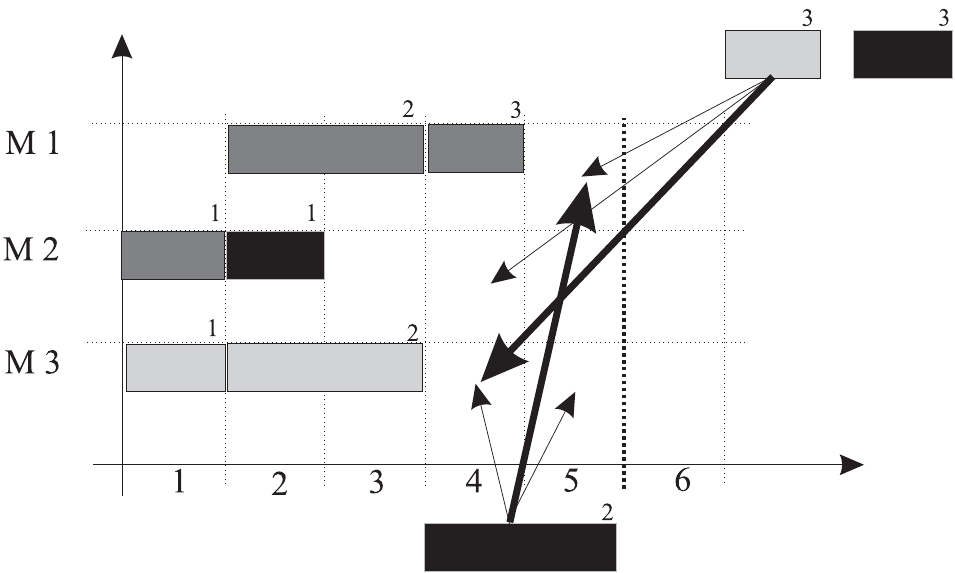}}
\subfigure[]{\includegraphics[width=.23\textwidth]{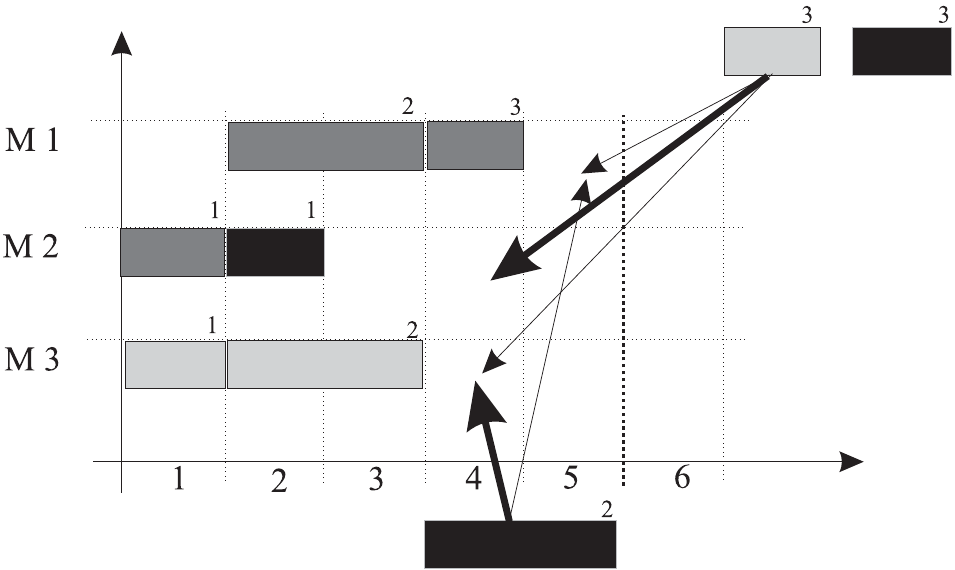}}
\subfigure[]{\includegraphics[width=.23\textwidth]{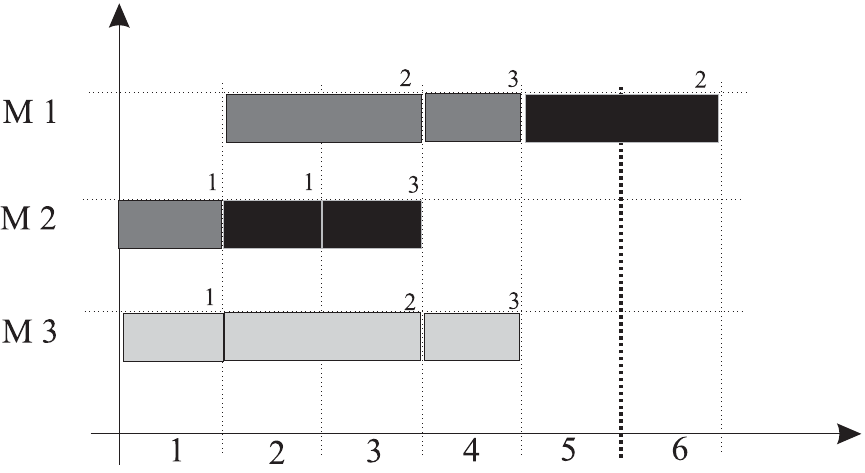}}
\subfigure[]{\includegraphics[width=.23\textwidth]{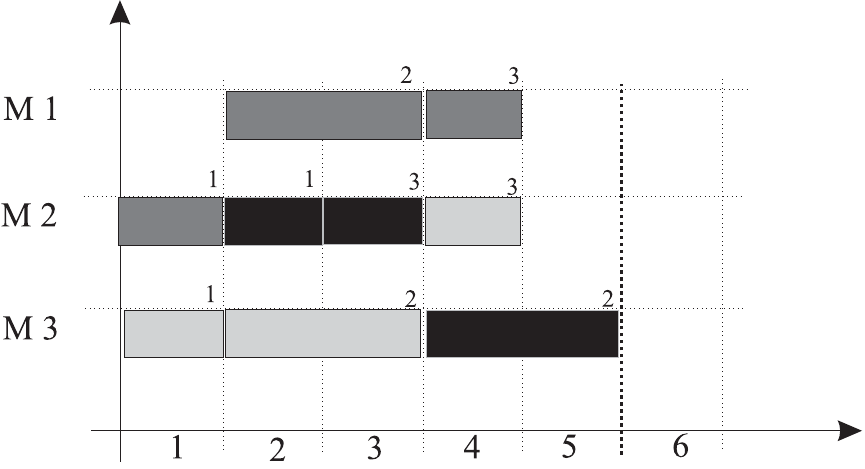}}
\caption{\em \small Example of the plan making for the minimal time and minimal transportation cost. There are three pieces to be manufactured ($P_1$ grey, $P_2$ white, $P_3$ black) with three working steps (with length 1, 2, 1). The $WS_2$ cannot be processed on M2. {\bf (a)} Decisions $WS_2(P_3)$ $\rightarrow$ M1 in position 5,  $WS_3(P_2)$  $\rightarrow$ M3 in position 4; {\bf (b)} Decisions $WS_2(P_3)$  $\rightarrow$ M3 in position 4,  $WS_3(P_2)$ $\rightarrow$ M2
in position 4; {\bf (c)} Final plan of (a); {\bf (d)} Final plan of (b).}
\label{beleg_plan_decisions}
\end{figure}

The first feature of agent-based optimization consists in a local
character of used data. Combinatorial or heuristic approaches
assume the data, required to be optimized, are globally available.
Optimization in this case looks like a "chess play" where all
pieces are visible and it needs to find some combination of pieces
positions. In the multi-agent variation there is no this central
viewpoint, each agent makes only a local decision about to occupy
a positions or not (see Fig.~\ref{beleg_plan_decisions}). In this
way the agent performs optimization of local decisions instead of
global positions of the working steps. Moreover, from agents
viewpoint, any of their decisions has not foreseeable perspectives
for a global optimum.

\begin{figure}[htp]
\centering
\includegraphics[width=.47\textwidth]{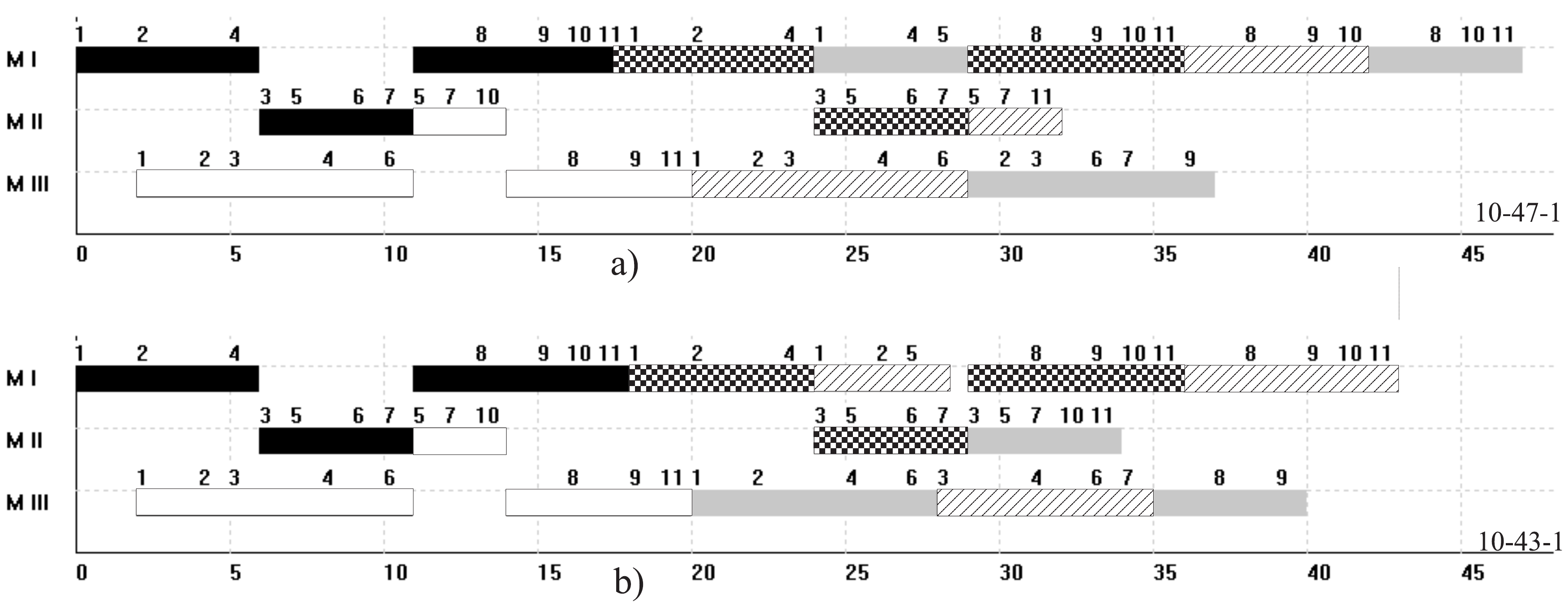}
\caption{\em \small The "forecasting" effect in assignment planning.
{\bf (a)} Each working step begins immediately after the previous step. The length of a whole plan is equal to 47;
{\bf (b)} The start of the working step 8 (3th piece) is delayed on one step, that allows reducing the common
length to 43 steps.}
\label{eigen_plans0}
\end{figure}

\begin{figure}[htp]
\centering
\includegraphics[width=.47\textwidth]{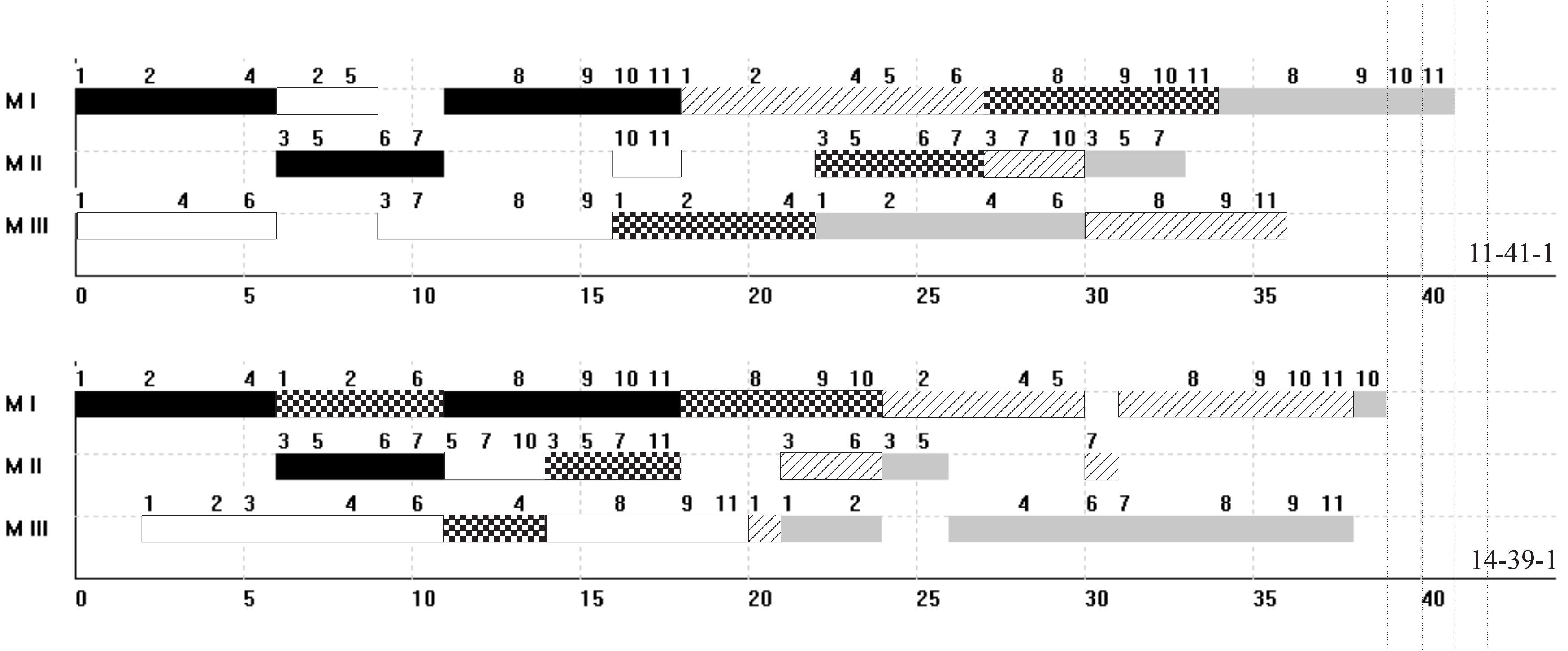}
\caption{\em \small Non optimal assignment plans with different number of jumps that result in different length
and transportation costs.}
\label{jumps}
\end{figure}

The second feature of agent-based optimization is caused by the
local nature of optimization problem. Each agent during the CSP
phase tries to occupy a position immediately after the previous
working step. This strategy is motivated from the manufacturing
side in trying to avoid a waiting time at processing elements
(machines). Evidently, this strategy cannot guarantee a global
optimum. Therefore the agent has to compute what will happen if
the next processing step will begin not immediately after the
previous step. It can be achieved by shifting a manufacturing of a
workpiece on some steps that increase a local cost of a plan (e.g.
intermediate storage) but reduce global costs (see
Figs.~\ref{eigen_plans0},~\ref{jumps}). This approach
(forecasting) is similar to a decision tree in distributed form.
As known, an increasing of the depth of tree rapidly increases the
search space.

After discussing the features of agents-based optimization, one can focus on the problem of assignment plan. There are two important steps, that the optimization needs to be performed on. Firstly, an order of the working steps in the group 2 and 4 (see Fig.~\ref{constraint}). Forasmuch as there are only 2881 combination between $WS_1$-$WS_{11}$, this optimization step can be performed by exhaustive search. The second point of optimization concerns the local decisions (concerning machine and position) made by agents. However search
space (taking into account the forecasting effect) grows in this case exponentially and e.g. even for 22 agent (2 production workpieces, forecast for next 5 positions) comes to $\sim 10^{10}$. Therefore exhaustive search methods like constraints optimization are inefficient even on very fast computer. The search
space can be essentially reduced if to take into account the following observation.

The assignment planning for different workpieces represents an iterative process where all iterations are very similar to one another. In this way the whole assignment plan represents a periodic pattern, that can be observed in Fig.~\ref{eigen_plans0}. Here there are two main patterns shown by black and white colors (order of the working steps as well as their positions on machines) that however differ in the last workpieces. It means that in case the optimal (or near optimal) scheme for the first iteration is found, next iteration can use the same scheme. The distributed approach being able to treat this kind of pattern-like problem is known as the ant colony optimization algorithm (ACO) \cite{Corne}.
This method originated from observation of ants in the colony. Going from the nest to the food source, every ant deposits a chemical substance, called pheromone, on the ground. In the decision point (intersection between branches of a
path) ants make a local decision based on the amount of pheromone, where a higher concentration means a shorter path. The result of this strategy represents a pattern of routes where thick line points to a shorter path. Similar strategy can be applied to local decisions of agents, participating in the plan making.

Agents after the CSP approach choose several assignment plans from the generated set of them and form an optimization pool. These assignment plans can represent also only segments of plans (these connected working steps represent independent parts of assignment plan) that satisfy all formulated constraints. These segments/plans can be combined into a common plan so that to satisfy the postulated optimization criterion. Thus, the more optimal segments are
included into this pool, the more optimal common plan will be obtained. The ACO algorithm marks (like a pheromone rate) the optimal segments obtained on the previous step. The fragments with the highest pheromone rate are included into
the top of pool. In this way agents consider first the ACO-obtained sequence and try to modify it (e.g. using forecasting effect). Thus, an optimization pool has always solutions with a high pheromone rate, from them the most optimal one will be then chosen.

\begin{figure}[htp]
\centering 
\includegraphics[width=.3\textwidth]{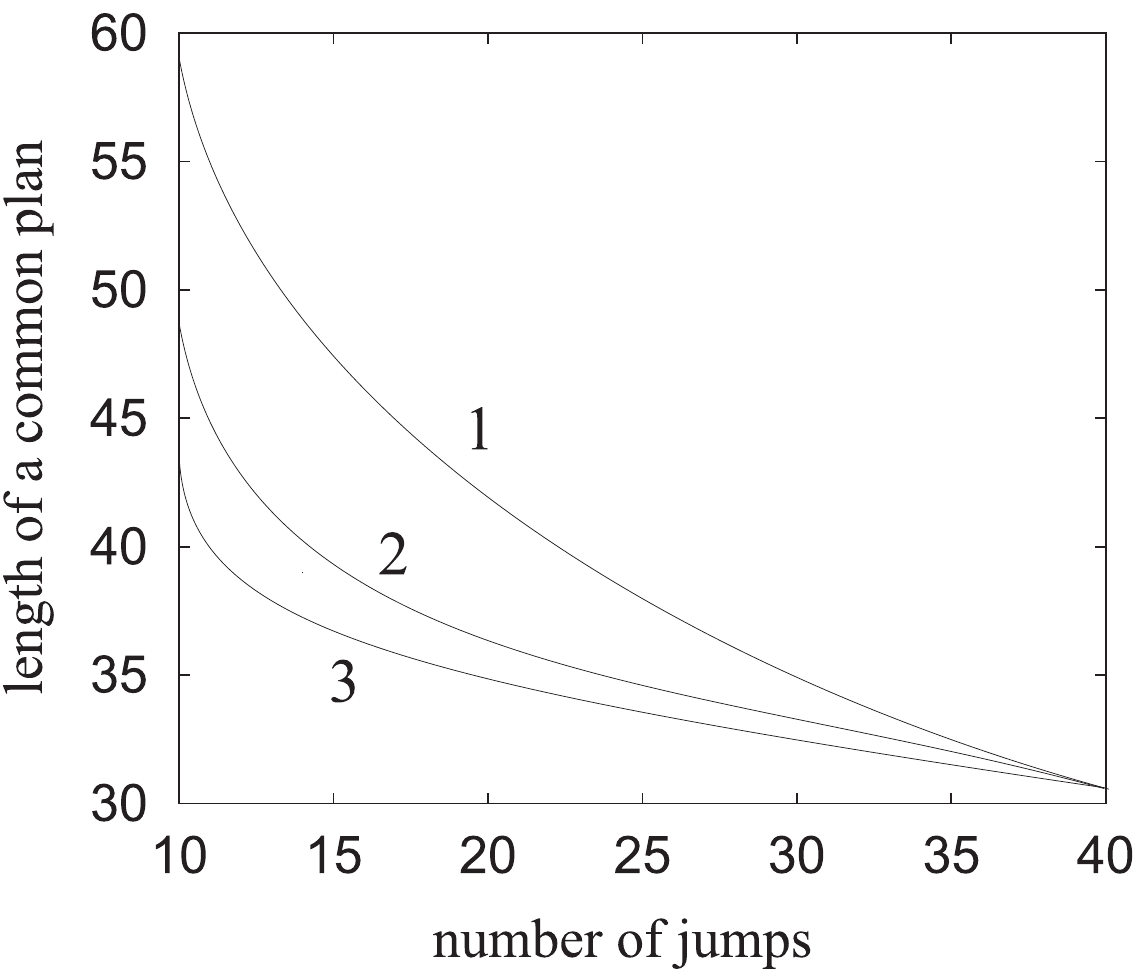}
\caption{\em \small Dependence between the number of jumps and the
length of generated plans. The curves 1, 2, 3 represent
the cases without forecasting, with one-position
forecasting for only the first workpiece, and with two-positions
forecasting for all workpieces.} \label{optim_plot}
\end{figure}

The optimality of a plan is also influenced by a number of
transportations of a workpiece from one machine to another. These
transportations are represented by so-called "jumps" in the plan
making, as shown in Fig.~\ref{jumps}. The minimal number of jumps
for a workpiece is defined by technological requirements and e.g.
for a plan shown in Fig.~\ref{eigen_plans0} is equal to 2. However
the number of jumps can be increased that worsens a cost but
improves other characteristics of an assignment plan. This
mechanism is utilized in combined optimization criteria, e.g. the
minimal cost at defined length (constant delivery date). Dependence between the number of jumps and, for example, the length of generated plans is shown in Fig.~\ref{optim_plot}.

\section{Conclusion}

The presented approach enables to react reasonably to disturbances
in manufacturing by using the constraint-based approach in a
multi-agent way. It does not require any centralized elements,
that essentially increases a reliability of common system.

\small

\end{document}